\documentclass{article}
\usepackage[top=120pt,bottom=120pt,left=108pt,right=108pt]{geometry}
\usepackage[utf8]{inputenc}
\usepackage{amsmath}
\usepackage{amsfonts}
\usepackage{graphicx}
\usepackage[numbers]{natbib}
\usepackage{algorithm}
\usepackage{algpseudocode}
\usepackage{multirow}
\usepackage{booktabs}
\usepackage{authblk}
\usepackage{enumitem}
\title{Percentile-based probabilistic optimization\\ for systematic and random uncertainties\\ in radiation therapy}

\author[1]{Albin Fredriksson}
\author[1]{Erik Engwall}
\author[2,3]{Jenneke de Jong}
\author[1]{Johan Sundström}
\affil[1]{RaySearch Laboratories, Stockholm, Sweden}
\affil[2]{Radiotherapy, Erasmus MC Cancer Institute, Rotterdam, Netherlands}
\affil[3]{Medical Physics and Informatics, HollandPTC, Delft, Netherlands}
\date{December 2025}

\begin{document}

\maketitle

\begin{abstract}

\noindent
Geometric uncertainty can degrade treatment quality in radiation therapy. While margins and robust optimization mitigate these effects, they provide only implicit control over clinical goal fulfillment probability. We therefore develop a probabilistic planning framework using a percentile-based optimization function that targets a specified probability of clinical goal fulfillment.

Systematic and random uncertainties were explicitly modeled over full treatment courses. A scenario dose approximation method based on interpolation between a fixed set of doses was used, enabling efficient simulation of treatment courses during optimization. The framework was evaluated on a prostate case treated with volumetric-modulated arc therapy (VMAT) and a brain case treated with pencil beam scanning (PBS) proton therapy. Plans were compared to conventional margin-based and worst-case robust optimization using probabilistic evaluation.

For the prostate case, probabilistic optimization improved organ at risk (OAR) sparing while maintaining target coverage compared to margin-based planning, increasing average OAR goal fulfillment probability by 13.3 percentage points and reducing 90th percentile OAR doses by an average of 3.5~Gy. For the brain case, probabilistic optimization improved target minimum dose passing probabilities (e.g., 88\% vs.~22\% for $D_{95}$) and brainstem maximum dose passing probability (70\% vs.~30\%), while maintaining comparable or improved OAR sparing compared to worst-case optimization.

Probabilistic optimization enables explicit and interpretable control over goal fulfillment probabilities. Combining full treatment course modeling with efficient approximate dose calculation, the proposed framework improved the trade-off between target coverage and OAR sparing compared to conventional planning approaches in both photon and proton therapy.

\end{abstract}

\section{Introduction}
Geometric uncertainty, such as patient setup and range uncertainty, can degrade radiation therapy treatment quality. Robust optimization has become the standard approach to mitigate these effects in proton therapy, and is increasingly being adopted for photon therapy when margins fail to provide adequate protection against non-rigid dose deformations~\cite{sterpin2024robustness, kaplan2022plan, archibald2017robust, zhang2018robust}.

Margin recipes often aim for a specified probability of target coverage under systematic and random setup errors~\cite{van2000probability}. Conventional robust optimization frameworks~\cite{fredriksson2012characterization, unkelbach2009reducing} enable implicit control over the probability through scenario selection---for example, the magnitudes of uncertainties to consider can be selected in accordance with the margin recipe~\cite{sterpin2024robustness}. However, neither margin-based planning nor conventional robust optimization explicitly optimizes the probability of goal fulfillment.

Probabilistic planning has been explored to provide such control. Löf et al.~\cite{lof1995optimal} simulated full treatment courses under systematic and random errors for 1D phantoms and maximized the probability of complication-free tumor control. Witte et al.~\cite{witte2007imrt}, Gordon et al.~\cite{gordon2010coverage} and Bohoslavsky et al.~\cite{bohoslavsky2013probabilistic} extended these ideas to clinical cases, using sampling to handle systematic errors and dose or fluence blurring for random errors. Witte et al.~focused on tumor control and normal tissue complication probability, whereas Gordon et al.~and Bohoslavsky et al.~optimized for a selected probability of goal fulfillment. Tilly et al.~\cite{tilly2019probabilistic} optimized the conditional value-at-risk (CVaR), iteratively adjusting its quantile parameter until the desired goal fulfillment probability was achieved.

Recent studies have considered full treatment course simulation for probabilistic evaluation~\cite{tilly2015fast, souris2019monte, rojo2023ptv, de2025probabilistic, olovsson2025robust}. In particular, Sterpin et al.~\cite{sterpin2021development} emphasize the importance of statistically sound robustness evaluation, showing that full Monte Carlo simulation of treatment courses can enable more accurate trade-offs than conventional evaluation strategies that use only a few scenarios. They also highlight the need for fast treatment course dose calculation to make comprehensive robustness evaluation practical. Examples of such fast approaches include methods that precompute a set of dose distributions and combine them by interpolation to simulate full treatment courses~\cite{rojo2023ptv, de2025sparse}.

Building on the work by Bohoslavsky et al.~\cite{bohoslavsky2013probabilistic}, we propose a probabilistic optimization approach that explicitly models both systematic and random uncertainties over full treatment courses. The framework directly targets a specified probability of meeting the planning goals using percentile-based optimization functions. To make the optimization computationally tractable, we adapt a recently proposed fast robustness evaluation method~\cite{de2025sparse} to the optimization setting, allowing efficient simulation of full treatment courses. We demonstrate the approach on a prostate case treated with volumetric-modulated arc therapy (VMAT) and a brain case treated with pencil beam scanning (PBS) proton therapy.

%The main contributions of the present work are the following:
%\begin{itemize}
%    \item Systematic and random errors are explicitly modeled
%    \item A probabilistic objective that allows for scaling between conditional expected value optimization and percentile optimization is formulated
%    \item Correlation between probabilistic goals is considered
%    \item A fast dose calculation method that can be used in the optimization is introduced
%    \item The probabilistic planning method is evaluated using both photon and proton therapy
%    \item The explicit modeling makes the method useful in many situations (although not tested here):
%    \begin{itemize}
%        \item For hypofractionated as well as conventionally fractionated treatments
%        \item It also allows probabilistic planning taking nonlinear effects into account, e.g., accumulating dose using the LQ model
%    \end{itemize}
%\end{itemize}

\section{Methods}
\begin{figure*}[htbp]
  \centering
  \includegraphics[width=\textwidth]{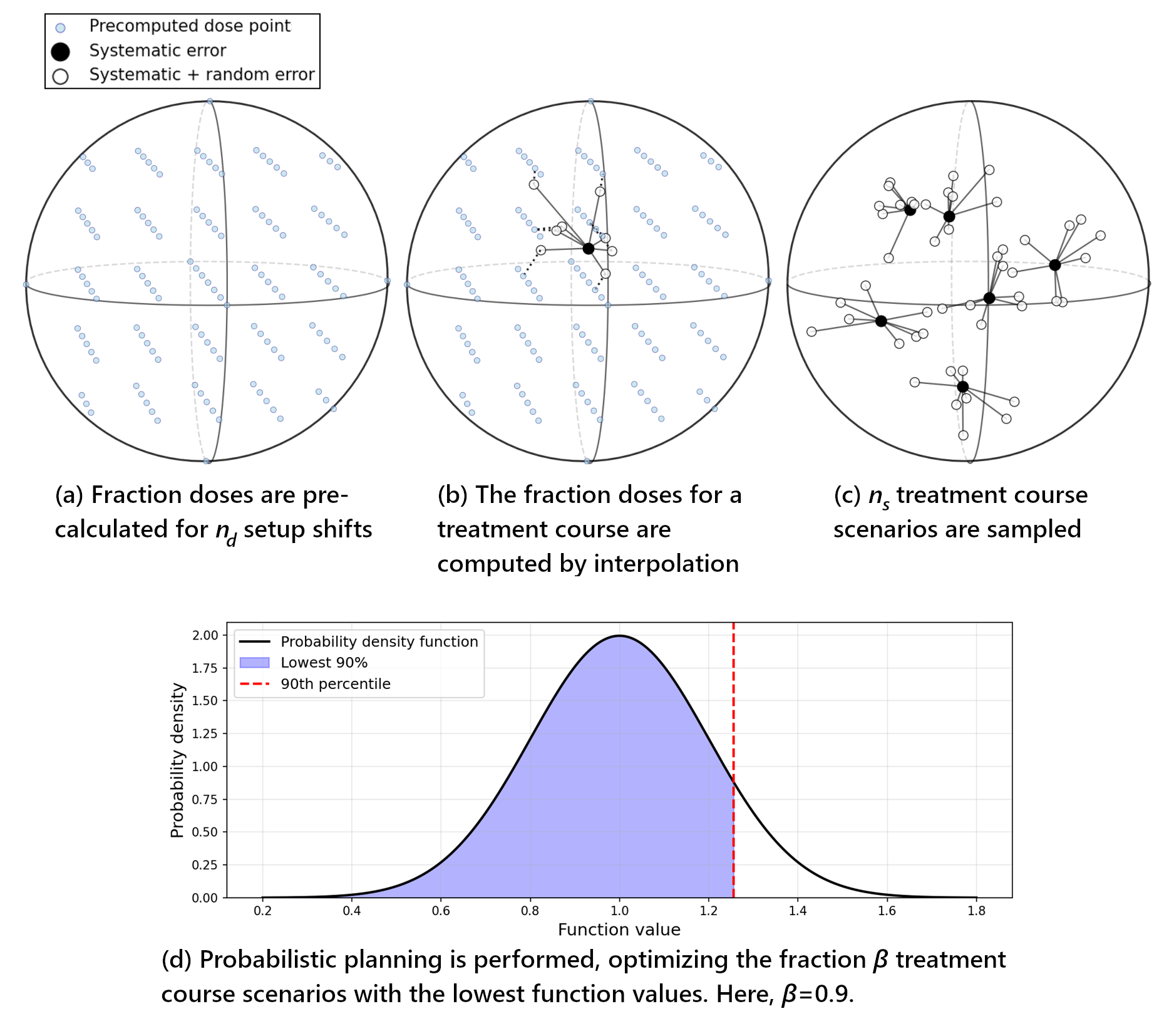}
  \caption{
    Overview of the proposed probabilistic optimization framework with approximate treatment course scenario dose calculation. (a) Fraction doses are precalculated on a grid of setup shifts. (b) A treatment course scenario is generated by sampling a systematic setup error and combining it with per-fraction random errors, and its constituent fraction doses are computed by interpolation in the grid (dotted lines indicate the nearest neighbors). (c) Multiple treatment course scenarios are sampled to form an empirical distribution of scenario doses. (d) Probabilistic optimization is performed using a percentile-based objective, targeting the fraction $\beta$ of best function values over the treatment course scenarios, i.e., the $100\beta$-percentile.
  }
  \label{fig:method}
\end{figure*}

An illustration of the overall framework for probabilistic optimization is given in Figure~\ref{fig:method}. The following subsections give a detailed description of its components.

\subsection{Treatment course scenarios}
We model a treatment course as a sequence of fractions affected by both systematic and random errors. Considering setup uncertainty only, a treatment course scenario (or just ``scenario'' for short) $s$ is represented by a setup shift for each fraction, composed of a systematic component shared across fractions and a random per-fraction component:
\[
s = (\Delta^{\textrm{syst}} + \Delta^{\textrm{rand}}_1,\ldots,\Delta^{\textrm{syst}} + \Delta^{\textrm{rand}}_{n_f})
\]
where $\Delta^{\textrm{syst}}$ denotes the systematic setup error, $\Delta^{\textrm{rand}}_i$ the random setup error at fraction $i$, and $n_f$ the number of fractions. The total dose $d(x; s)$ in scenario $s$ for variables $x$ (e.g., spot weights for protons and multi-leaf collimator leaf positions, dose rate, and gantry speed for photons) is given by
\[
d(x; s) = \sum_{i = 1}^{n_f} d_f(x; \Delta^{\textrm{syst}} + \Delta^{\textrm{rand}}_i)
\]
where $d_f(x; \Delta)$ is the dose distribution from one fraction evaluated for variables $x$ and setup error $\Delta$. The formulation can easily be extended to include other uncertainty sources, such as range uncertainty.

\subsection{Percentile-based optimization functions}
To enable direct control over the probability of goal fulfillment, we introduce a \emph{percentile-trimmed power mean} that optimizes over the fraction $\beta$ best scenarios. A similar approach has been described by Bohoslavsky et al.~\cite{bohoslavsky2013probabilistic}, who optimized the expected value over a fraction of the best scenarios. Here, we generalize their formulation to obtain a measure that comes closer to the percentile (value-at-risk).

Consider an optimization function $f(d(x;S))$ evaluated on the dose under scenario $S$, where the random variable $S$ takes the value $s$ from the set $\Omega$ of scenarios with probability $p_s$. Define the objective values for each scenario by
\[
v_s = f(d(x; s)), \qquad s \in \Omega.
\]
Order the values $v_{(1)} \leq \ldots \leq v_{(n_s)}$ with corresponding probabilities $p_{(i)}$, where $n_s$ is the number of scenarios in $\Omega$, and introduce cumulative probabilities
\[
c_j = \sum_{i=1}^j p_{(i)}, \qquad j = 1, \ldots, n_s, \qquad c_0 = 0.
\]
For a target percentile level $100\beta$, $\beta\in (0,1]$, let $k$ be the smallest index satisfying $c_{k-1} < \beta \leq c_k$. Because multiple scenarios may share the same function value, denote by
\[
p_{\textrm{tie}} = \sum_{i: v_{(i)} = v_{(k)}} p_{(i)}
\]
the probability mass of the tie group with function values equal to $v_{(k)}$. Let $t$ be the smallest index of that tie group.

From these quantities, we construct the trimmed distribution supported on the fraction $\beta$ best scenarios and renormalize to get the conditional probabilities
\[
\tilde p_{(i)} = \left\{
\begin{array}{ll}
\frac{p_{(i)}}{\beta}, \quad & v_{(i)} < v_{(k)},\\
\frac{p_{(i)}}{\beta} \frac{\beta - c_{t-1}}{p_{\textrm{tie}}}, &  v_{(i)} = v_{(k)},\\
0 & \textrm{otherwise.}
\end{array}
\right.
\]
The percentile-trimmed power mean with power $a \neq 0$ at percentile $100\beta$ is then
\[
Q_a^\beta(f(d(x;S))) = \left( \sum_{i=1}^{n_s} \tilde p_{(i)} v_{(i)}^a \right)^{1/a}.
\]

Intuitively, the percentile-trimmed power mean evaluates the function over the best-performing fraction of scenarios, smoothly emphasizing worst-case behavior within this subset as the power parameter increase.

Special cases:
\begin{itemize}
    \item $a=1$: the expected value conditioned on belonging to the best fraction $\beta$ of scenarios, as in Bohoslavsky et al.~\cite{bohoslavsky2013probabilistic}.
    \item $a \to \infty$: the $100\beta$-percentile (value-at-risk).
\end{itemize}
Intermediate values of $a$ provide a continuous trade-off between conditional expected value and percentile behavior.

For $a \neq 1$, the function $Q_a^\beta$ requires $f$ to be nonnegative. Other approximations of the percentile could be used to allow also functions $f$ taking negative values, e.g., taking the log-sum-exp over the conditional probability distribution $\tilde p$.

\subsection{Correlation}
Multiple optimization functions can either be handled independently or jointly. If two functions $f$ and $g$ are optimized using separate percentile functions,
\[
Q_a^\beta(f) + Q_a^\beta(g),
\]
then the optimization would aim for a plan where each function has a low value in a fraction $\beta$ of the scenarios, but this subset of scenarios might be different for the two functions.

Alternatively, optimizing the combined function
\[
Q_a^\beta(f + g).
\]
aims for a low value for the sum of the functions within the same subset of scenarios, effectively coupling their probabilities. Another option is to choose a dominant percentile function and reuse its scenario ordering for correlated percentile functions.

\subsection{Optimization formulation}
The overall optimization problem, taking correlation into account in the objective, is 
\begin{align*}
\underset{x \in X}{\text{minimize}} \quad & Q_a^\beta\left( \sum_{i \in O_\textrm{prob}} w_i f_i(d(x; S))\right) + \sum_{i \in O_\textrm{nom}} w_i f_i(d(x;s_{\textrm{nom}})\\
\text{subject to} \quad
&  \begin{array}{lrll}
        Q_a^\beta(f_i(d(x; S))) &\leq & 0, & i \in C_\textrm{prob},\\
        f_i(d(x; s_{\textrm{nom}})) &\leq& 0, & i \in C_\textrm{nom},\\
    \end{array}
\end{align*}
where $x$ denotes the variables, $X$ is the set of feasible variable values, $O_\textrm{prob}$ and $O_\textrm{nom}$ are sets indexing the objective functions to be treated, respectively, probabilistically and nominally, and $C_\textrm{prob}$ and $C_\textrm{nom}$ are similar but for constraints, $w_i$ is the importance weight of function $f_i$ for $i \in O_\textrm{prob} \cup O_\textrm{nom}$, and $s_{\textrm{nom}}$ denotes the nominal scenario. In this formulation, correlation is only taken into account in the objective and not in the constraints. Such correlation could however be achieved by letting the percentile objective function dominate the scenario selection for all percentile constraint functions.

\subsection{Treatment course scenario dose approximation}\label{sec-approx}
Simulating full treatment courses involves many dose calculations: for $n_f$ fractions and $n_s$ scenarios, $n_f \times n_s$ dose calculations are required per iteration. To reduce computational cost, we adopt the interpolation strategy from Fredriksson et al.~\cite{fredriksson2021robust}, extended to handle both systematic and random errors by De Jong et al.~\cite{de2025sparse}. This approach drastically reduces the computational burden and has been shown to preserve adequate accuracy for robustness evaluation~\cite{de2025sparse}.

In each iteration, for the current optimization variables $x$, we precompute dose distributions $\hat d_j(x)$, $j=1,\ldots,n_d$, on a regular 3D grid for setup shifts that are at most a distance $R$ from the origin, with $R$ selected to cover most of the uncertainty. The fraction dose $d_f(x;\Delta)$ for an arbitrary shift $\Delta$ is then approximated by
\[
d_f(x;\Delta) = \sum_{j=1}^{n_d} \alpha_j(\Delta) \hat d_j(x),
\]
where $\alpha_j(\Delta)$ are nonnegative interpolation weights, which can be selected using, e.g., nearest neighbor interpolation, inverse distance weighting, or trilinear interpolation.

The accumulated dose $d(s)$ over a treatment course is
\[
d(x; s) = \sum_{i=1}^{n_f} \sum_{j=1}^{n_d} \alpha_j(\Delta^{\text{syst}} + \Delta^{\text{rand}}_i) \hat d_j(x) = \sum_{j=1}^{n_d} \gamma_j(s) \hat d_j(x), \qquad  \gamma_j(s) = \sum_{i=1}^{n_f} \alpha_j(\Delta^{\text{syst}} + \Delta^{\text{rand}}_i).
\]
This reduces the number $n_d$ of dose calculations to the number of precomputed grid points. When the grid spacing is $R/2$, $R/3$, or $R/4$ and points outside the sphere of radius $R$ are omitted, $n_d$ becomes respectively $33$, $123$, or $257$. Computing treatment course scenario doses is then a matter of weighted summation of these doses.

\subsection{Computational study}
Probabilistic optimization using the percentile-trimmed power mean was implemented in a research version of RayStation v2025 (RaySearch Laboratories, Sweden). It was applied to two patient cases and compared to conventional planning: a prostate case treated with VMAT where the conventional planning used margins, and a brain case treated with PBS protons where the conventional planning used worst-case optimization considering systematic range and setup uncertainty.

In the experiments, we used $\beta = 0.9$, corresponding to 90\% goal fulfillment and $a=8$, which is the power used in RayStation's approximation of worst-case optimization. Systematic and random setup errors were sampled from zero-mean isotropic 3D Gaussian distributions with standard deviations $\Sigma$ and $\sigma$, respectively, and the systematic range uncertainty was sampled from a zero-mean 1D Gaussian with standard deviation $\Sigma_r$.

The VMAT optimization used a fast dose calculation algorithm based on singular value decomposition combined with collapsed cone computations at intermediate iterations, which were used to update the dose. The evaluation used collapsed cone calculations for each perturbed dose.

The PBS optimization used RayStation's approximate dose calculation, performing interpolation between the Monte Carlo calculated spots~\cite{janson2024treatment}, and the evaluation used full Monte Carlo calculation for each perturbed dose. A constant radiobiological effectiveness (RBE) factor of 1.1 was used.

The treatment course scenario dose approximation with spacing $R/2$ (33 grid points) was used in the optimization and $R/3$ (123 grid points) in the evaluation, and $R$ was set to $3\sqrt{\Sigma^2 + \sigma^2}$, based on the settings found to provide adequate accuracy in De Jong et al.~\cite{de2025sparse}. For the probabilistic PBS plan, range  uncertainty was considered. To this end, the interval $[-2.6\Sigma_r, 2.6\Sigma_r]$ was discretized into 5 points in the optimization and 7 points in the evaluation, for a total of respectively $5 \times 33 = 165$ and $7 \times 123 = 861$ dose calculation points, which could be compared to the conventional robust optimization using $3 \times 7 = 21$ dose calculation points. The interpolation weights $\alpha$ were selected by nearest neighbor interpolation. In optimization as well as evaluation, $m=1000$ treatment courses were simulated, but using different random seeds and hence different treatment courses.

\subsection{Patient cases}
Specifics for the different cases were as follows:

\paragraph{Prostate case}
For the prostate case treated with VMAT, $\Sigma = 0.2$ cm and $\sigma = 0.3$ cm was used. The conventional plan used a margin of $0.7$ cm according to the Van Herk rule ($2.5 \Sigma + 0.7 \sigma$)~\cite{van2000probability}. The prescription was $77$ Gy to the clinical target volume (CTV) delivered in $35$ fractions.

The probabilistic optimization used a probabilistic constraint for the CTV prescription, and probabilistic objectives for the organ at risk (OAR) clinical goals, as well as nominal objectives promoting a sharp dose falloff. The conventional planning used the same optimization formulation, but with nominal functions instead of probabilistic ones, and with the margin-expanded target substituted for the CTV.

\paragraph{Brain case}
For the brain case treated with PBS, $\Sigma = 0.07$ cm and $\sigma = 0.2$ cm was used, as well as $\Sigma_r = 1.53$\%. The conventional plan considered systematic setup uncertainty of $0.34$ cm, based on the margin recipe for 95\% coverage probability ($2.79 \Sigma + 0.7 \sigma$)~\cite{van2000probability}, as well as systematic range uncertainty of $1.96\Sigma_r = 3$\% to also cover $95$\% of the range uncertainty and hence a total of $95\% \times 95\% \approx 90\%$ of the uncertainty space. The prescription was 60 Gy (RBE) delivered to the CTV in 30 fractions. However, as the CTV was directly adjacent to the brainstem, the main target clinical goal was relaxed to $D_{95} \geq 57$ Gy (RBE).

In the optimization, the probabilistic method used probabilistic constraints for the ``at most'' clinical goals for the OARs and CTV, and probabilistic objectives for the CTV minimum dose clinical goals, as well as nominal objectives promoting a sharp dose falloff and low dose to the brain. The conventional method used the same optimization formulation, but with worst-case robust functions instead of probabilistic ones.

\section{Results}

\subsection{Prostate case (VMAT)}
Probabilistic optimization consistently improved OAR sparing while maintaining target coverage (Table~\ref{tab:prostate}). Over the ten OAR goals, the probabilistic plan increased the average fulfillment probability by $13.3$ percentage points and reduced the P90 (90th percentile) OAR dose metrics by an average of $3.5$ Gy compared to the conventional plan. The P90 of the mean external dose was reduced by $0.2$\ Gy.

\begin{figure*}[htbp]
  \centering
  \includegraphics[width=\textwidth]{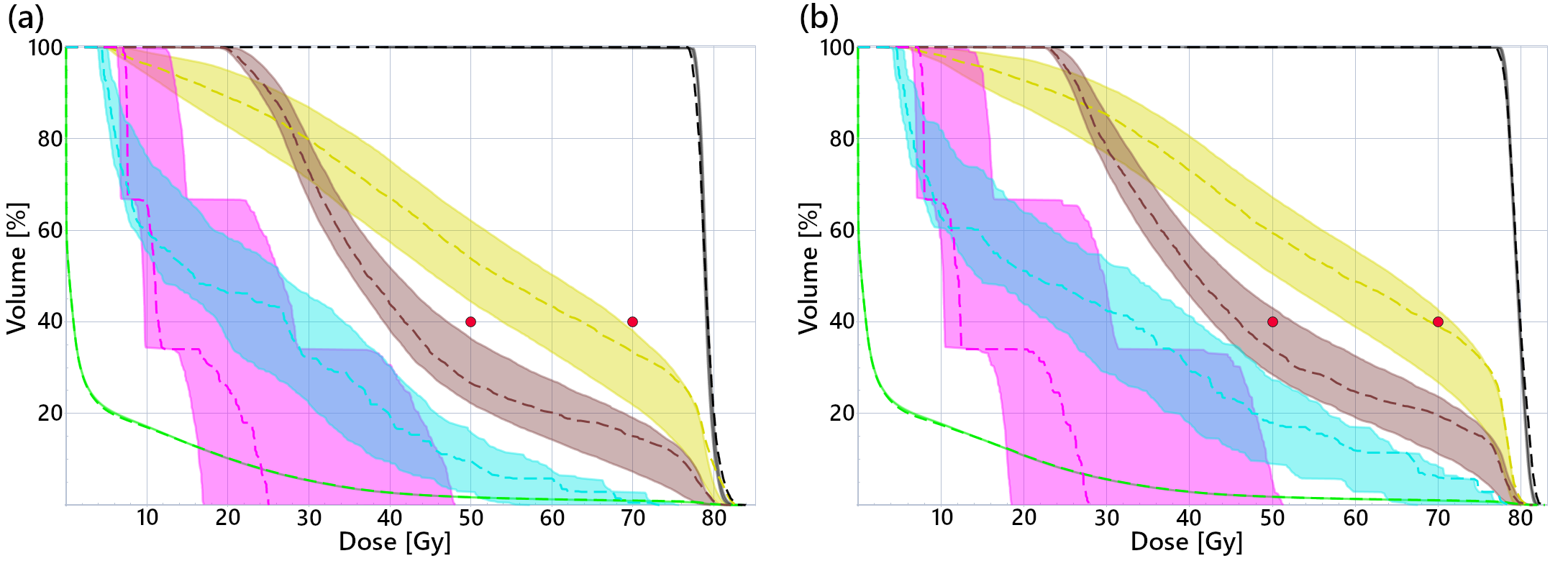}
  \caption{DVHs and 10--90th percentile DVH bands for the prostate case treated with VMAT for (a) the probabilistic method and (b) the conventional margin-based method. ROIs displayed are CTV (black), rectum (brown), bladder (yellow), anal canal (cyan), bulbus (magenta), external (green). Nominal DVHs are shown as dashed lines. Two points at (50\ Gy,~40\%) and (70\ Gy, 40\%) have been added to both graphs for reference.}
  \label{fig:prostate-dvh}
\end{figure*}

\begin{figure}[htbp]
  \centering
  \includegraphics[width=\textwidth]{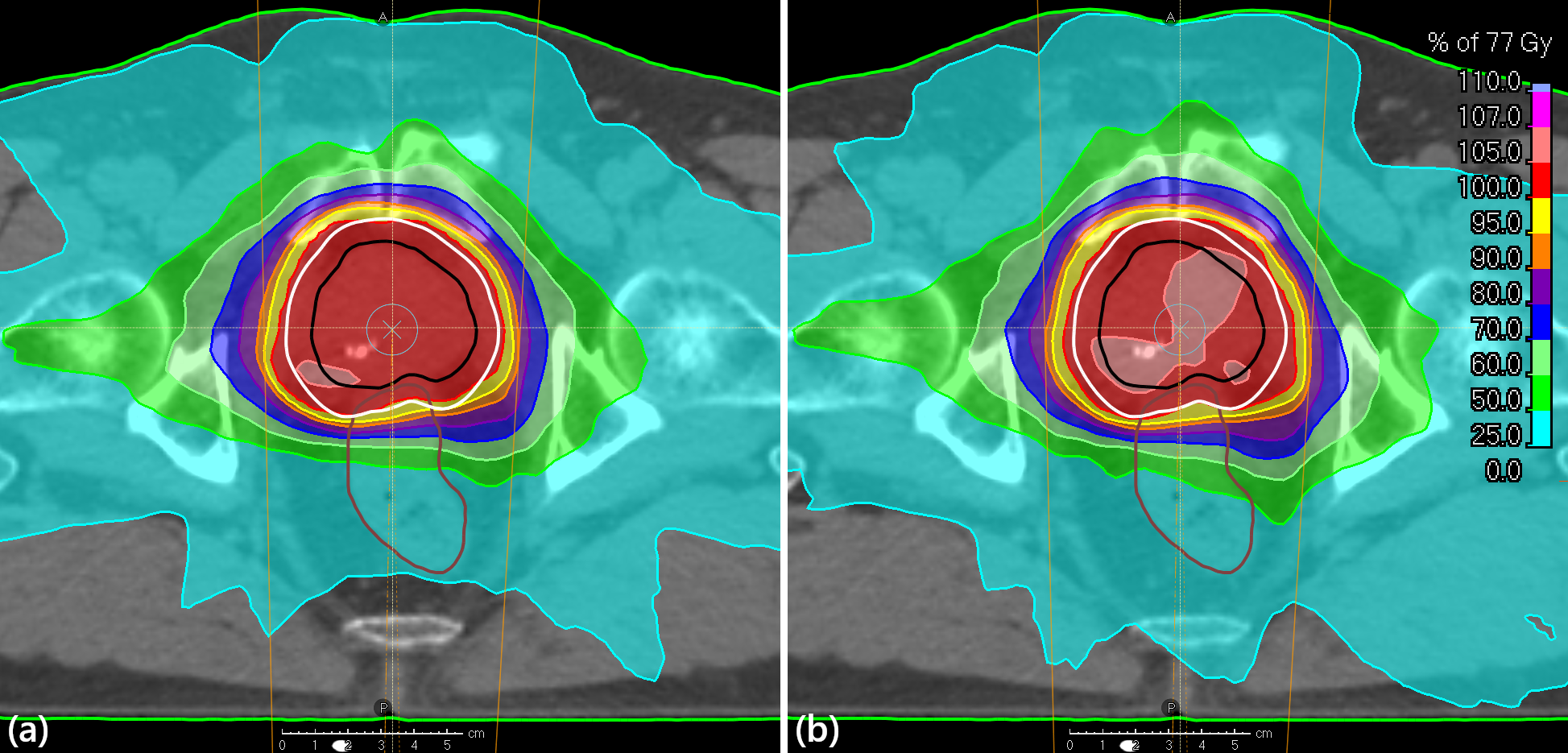}
  \caption{Transversal slices and dose distributions of the prostate case treated with VMAT for (a) the probabilistic method and (b) the conventional margin-based method. ROIs displayed are CTV (black), CTV + 0.7 cm margin (white), rectum (brown), bladder (yellow), anal canal (cyan), bulbus (magenta), external (green).}
  \label{fig:prostate-dose}
\end{figure}

The dose--volume histograms (DVHs) in Figure~\ref{fig:prostate-dvh} illustrate these differences, showing lower OAR doses and maintained target coverage in the probabilistic plan and the dose distributions in Figure~\ref{fig:prostate-dose} show that it has slightly retracted the high-dose region away from the rectum.

\begin{table}[ht]
\centering
\caption{Probabilistic evaluation statistics for the prostate case treated with VMAT: clinical goal fulfillment rates and percentile dose statistic values over the evaluation scenarios for the probabilistic and conventional margin-based optimization. For ``at least'' goals, the 10th percentile (P10) values are shown, and for ``at most'' goals, the 90th percentile (P90) values.}
\label{tab:prostate}
\vspace{0.25cm}
\scalebox{0.96}{
\begin{tabular}{llrrrr}
\hline
\textbf{ROI} & \textbf{Clinical goal} &
\multicolumn{2}{c}{\textbf{Probabilistic}} &
\multicolumn{2}{c}{\textbf{Conventional}} \\
& &
\textbf{Passed} & \textbf{P10 or P90} & \textbf{Passed} & \textbf{P10 or P90} \\
&& \textbf{\scriptsize (\%)} & \textbf{\scriptsize (Gy)} & \textbf{\scriptsize (\%)} & \textbf{\scriptsize (Gy)} \\
\hline

\multirow{3}{*}{CTV}
& $D_{100} \ge 77.0\ \text{Gy}$ & 72 & 76.1 & 71 & 76.4 \\
& $D_{99.8} \ge 76.9\ \text{Gy}$ & 97 & 77.2 & 99 & 77.4 \\
& $D_{99} \ge 76.2\ \text{Gy}$ & 100 & 77.4& 100 & 77.6\\
& $D_{2} \le 82.4\ \text{Gy}$ & 100 & 81.4& 100 & 81.6\\
\addlinespace
\multirow{4}{*}{Rectum}
& $D_{75} \le 28.6\ \text{Gy}$ & 23 & 31.3 & 1 &  33.6\\
& $D_{50} \le 42.9\ \text{Gy}$ & 98 & 41.0 & 72 & 45.0\\
& $D_{20} \le 66.7\ \text{Gy}$ & 80 & 69.0 & 50 & 73.2\\
& $\bar D$ & -- & 46.8 & -- & 46.8 \\
\addlinespace
\multirow{4}{*}{Bladder}
& $D_{50} \le 61.9\ \text{Gy}$ & 93 & 60.3 & 78 & 64.7\\
& $D_{25} \le 71.4\ \text{Gy}$ & 32 & 76.8 & 11 & 77.2\\
& $D_{10} \le 72.0\ \text{Gy}$ & 0 & 79.2 & 0 & 78.7\\
& $\bar D$ & -- & 56.5 & -- & 56.5 \\
\addlinespace
\multirow{3}{*}{Anal canal}
& $D_{60} \le 33.3\ \text{Gy}$ & 100 & 18.3 & 100 & 23.5\\
& $D_{40} \le 38.1\ \text{Gy}$ & 99 & 34.2 & 84 & 40.7\\
& $\bar D \le 45.0\ \text{Gy}$ & 100 & 28.9 & 100 & 34.4\\
\addlinespace
Bulbus & $\bar D \le 30.0\ \text{Gy}$ & 93 & 28.3& 89 & 31.3\\
\addlinespace
External & $\bar D$ & -- & 5.5& -- & 5.7\\
\hline
\end{tabular}
}
\end{table}

\subsection{Brain case (PBS)}
The probabilistic plan improved target coverage while maintaining or improving OAR sparing, increasing the passing probability for maximum dose to brainstem core from $30$\% to $70$\% (Table~\ref{tab:brain}). For the CTV, the fulfillment probabilities for the $D_{95}$ and $D_{98}$ goals were $88$\% and $29$\%, respectively, compared to $22$\% and $5$\% for the conventional method. On the other hand, the passing probability for the CTV maximum goal was $5$ percentage points lower for the probabilistic approach. The probabilistic method lead to $1.6$ Gy (RBE) lower P90 levels for both the chiasm and optic nerve goals, while the brainstem goal levels were similar. It also lead to P90 reductions of $0.6$ Gy (RBE) and $0.1$ Gy (RBE) to the mean brain and external doses, respectively.

\begin{figure}[htbp]
  \centering
  \includegraphics[width=\textwidth]{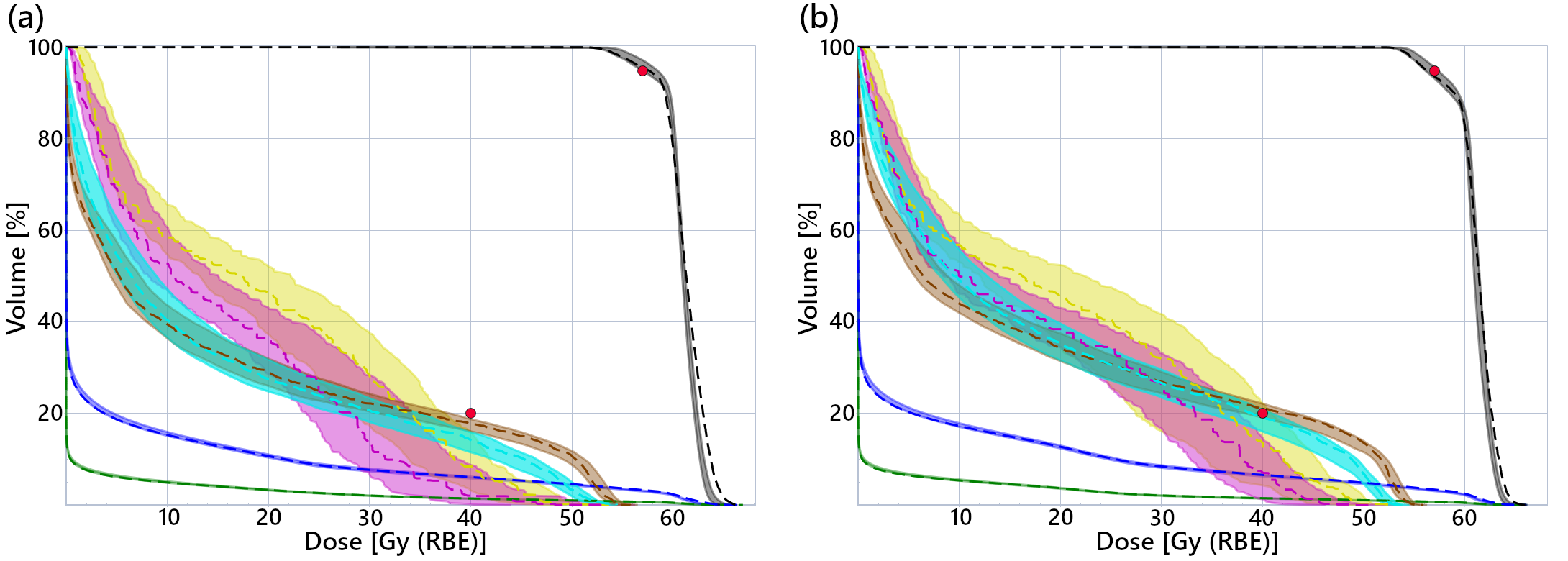}
  \caption{DVHs and 10--90th percentile DVH bands for the brain case treated with PBS for (a) the probabilistic method and (b) the conventional robust optimization. ROIs displayed are CTV (black), brainstem core (cyan), brainstem surface (brown), optic nerve left (yellow), chiasm (magenta), brain (blue), external (green). Nominal DVHs are shown as dashed lines. Two points at (40~Gy,~20\%) and (57\ Gy, 95\%) have been added to both graphs for reference.}
  \label{fig:brain-dvh}
\end{figure}

\begin{figure}[htbp]
  \centering
  \includegraphics[width=\textwidth]{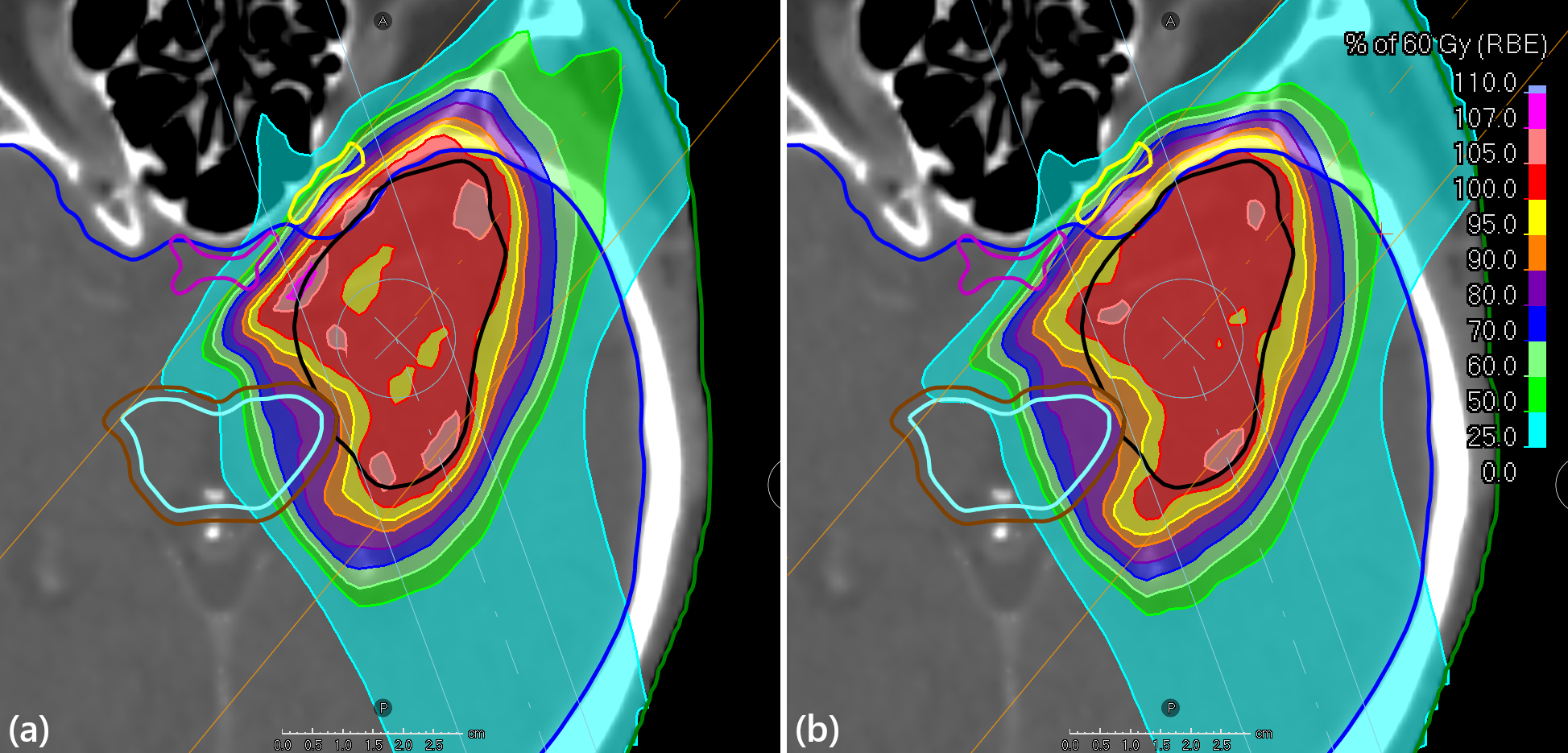}
  \caption{Transversal slices and dose washes of the brain case treated with PBS for (a) the probabilistic method and (b) the conventional robust optimization. ROIs displayed are CTV (black), brainstem core (cyan), brainstem surface (brown), optic nerve left (yellow), chiasm (magenta), brain (blue), external (green).}
  \label{fig:brain-dose}
\end{figure}

Figure~\ref{fig:brain-dvh} shows that the DVHs of the OARs of the probabilistic plan were generally lower than those of the conventional plan, and that the CTV coverage was slightly higher. The dose distributions in Figure~\ref{fig:brain-dose} show that the probabilistic plan retracted the dose away from the inside of the brainstem but had higher doses just outside of it than the conventional plan. It also used the anterior beam more than the conventional plan, and had more heterogeneous nominal target dose.

\begin{table}[ht]
\centering
\caption{Probabilistic evaluation statistics for the brain case treated with PBS: clinical goal fulfillment rates and percentile dose statistic values over the evaluation scenarios for the probabilistic and conventional robust optimization. All dose statistics are presented in RBE-weighted dose. For ``at least'' goals, the 10th percentile (P10) values are shown, and for ``at most'' goals, the 90th percentile (P90) values.}
\label{tab:brain}
\vspace{0.25cm}
\scalebox{0.96}{
\begin{tabular}{llrrrr}
\hline
\textbf{ROI} & \textbf{Clinical goal} &
\multicolumn{2}{c}{\textbf{Probabilistic}} &
\multicolumn{2}{c}{\textbf{Conventional}} \\
& & \textbf{Passed} &  \textbf{P10 or P90} & \textbf{Passed} & \textbf{P10 or P90}\\
&& \textbf{\scriptsize (\%)} & \textbf{\scriptsize (Gy)} & \textbf{\scriptsize (\%)} & \textbf{\scriptsize (Gy)} \\
\hline

\multirow{3}{*}{CTV}
& $D_{98} \ge 55.8\ \text{Gy}$ & 29 & 54.5 & 5 & 54.5\\
& $D_{95} \ge 57.0\ \text{Gy}$ & 88 & 56.9 & 22 & 56.0\\
& $D_{2} \le 64.2\ \text{Gy}$ & 94 & 64.1 & 99 & 63.9\\
\addlinespace
\multirow{2}{*}{Brainstem surf.}
& $D_{0.03\,\text{cm}^3} \le 56.0\ \text{Gy}$ & 99 & 55.2 & 100 & 55.1\\
& $D_{0} \le 56.0\ \text{Gy}$ & 71 & 56.5 & 70 & 56.3\\
\addlinespace
\multirow{2}{*}{Brainstem core}
& $D_{0.03\,\text{cm}^3} \le 54.0\ \text{Gy}$ & 100 & 53.3 & 99 & 53.5\\
& $D_{0} \le 54.0\ \text{Gy}$ & 70 & 54.4 & 30 & 54.5\\
\addlinespace
Chiasm & $D_{0.03\,\text{cm}^3} \le 54.0\ \text{Gy}$ & 100 & 42.6 & 100 & 44.2\\
\addlinespace
Optic nerve (L) & $D_{0.03\,\text{cm}^3} \le 54.0\ \text{Gy}$ & 100 & 47.2 & 100 & 48.8\\
\addlinespace
Brain & $\bar D$ & -- & 6.0 & -- & 6.6\\
\addlinespace
External & $\bar D$ & -- & 1.8 & -- & 1.9 \\
\hline
\end{tabular}
}
\end{table}

\section{Discussion}
The percentile-based probabilistic optimization produced plans with higher probabilities of clinical goal fulfillment and improved trade-offs between target coverage and OAR sparing compared to conventional planning. These improvements arise from the method's explicit control of the distribution of scenario doses over full treatment courses.

While this work focuses on probabilistic optimization, a natural first step toward probabilistic planning is the adoption of probabilistic evaluation. Applying probabilistic evaluation to conventionally optimized plans already enables a more informative assessment of clinical goal fulfillment probabilities than nominal or worst-case analysis, and can be introduced independently of changes to the optimization~\cite{sterpin2021development, de2025probabilistic}.

A discrepancy remains between the probabilistic constraints used during optimization and the passing probabilities observed in evaluation. For example, in the prostate case, a constraint targeting a $90$\% probability of CTV minimum dose resulted in passing rates of 97\%  for $D_{99.8}$ and 72\% for $D_{100}$. Such deviations can result from limited numbers of simulated treatment courses, constraint tolerances, differences between optimization and evaluation dose calculation algorithms, and the sensitivity of binary evaluation criteria to small dose variations.

Compared to previous probabilistic planning methods, the present framework introduces two main methodological extensions. First, both systematic and random errors are explicitly modeled without relying on dose blurring, enabled by a fast scenario dose approximation method. This makes the framework applicable not only to photon treatments in approximately homogeneous media, but also to photon treatments with pronounced density heterogeneities and to ion therapy. Second, the proposed percentile-trimmed power mean formulation allows smooth scaling between conditional expected value and percentile-based optimization while handling correlation between probabilistic goals.

Explicit modeling of random errors over full treatment courses enables future investigations of fractionation-dependent effects. For example, applications in hypofractionated or adaptive treatments and incorporation of radiobiological models, such as the linear-quadratic model, into the dose accumulation could be explored. For the approximate treatment course scenario dose calculation, alternative grids and interpolation schemes could be studied. In photon therapy for cases with limited heterogeneity in patient density, substantial speed ups could be achieved using the static dose cloud approximation~\cite{witte2007imrt, bohoslavsky2013probabilistic}.

Percentile-based optimization is inherently non-convex and therefore does not guarantee convergence to a global optimum. Nonetheless, the percentile-trimmed power mean is closely related to DVH-based optimization functions, which have a long history of successful clinical use despite similar non-convexity. Convex surrogates such as CVaR generally do not yield optimal solutions to the percentile criterion, but can serve as initializations that are subsequently refined using the non-convex percentile formulation.

This study does not consider anatomical changes or distributional robustness. Simplified anatomical variations could be incorporated using interpolation-based approaches~\cite{fredriksson2021robust}. However, combining anatomical changes with setup errors substantially increases the computational complexity. Robustness to uncertainty misspecification could be improved by the inclusion of random errors with different distributional assumptions in both the optimization and evaluation.

\section{Conclusion}
We have presented a probabilistic optimization framework for robust radiation therapy treatment planning that explicitly controls the probability of goal fulfillment, and evaluated the framework using both VMAT and proton PBS planning. The percentile-trimmed power mean provides an interpretable risk measure spanning expected value conditioned on a percentile and direct percentile optimization. Combined with efficient approximate dose calculation, the method enables realistic simulation and optimization of fractionated treatment courses under uncertainty. The example cases show that the method can yield improved probabilistic trade-offs between target coverage and OAR sparing compared to conventional planning methods.

\end{document}